\pdfoutput=1
\documentclass{article}
\usepackage{arxiv}
\usepackage[utf8]{inputenc}
\usepackage[style=numeric-comp,sorting=none]{biblatex} %Imports biblatex package
\AtEveryBibitem{\clearfield{url}} % This will clear the URL field
\AtEveryBibitem{\clearfield{note}} % This will clear the note field, if the 
\AtEveryBibitem{\clearfield{urldate}}
\renewbibmacro{urldate}{\iffieldundef{url}{}{\printurldate}}

\usepackage{layouts}
\usepackage{tikz}
\usepackage{multirow}
\usetikzlibrary{patterns}
\usepackage[colorlinks=true,citecolor=blue,linkcolor=magenta]{hyperref}
\usepackage{graphicx}
\usepackage{pgf}
\usepackage{caption}
\usepackage{subcaption}
\usepackage{dcolumn}
\usepackage{bm}
\usepackage{stmaryrd}
\usepackage[bbgreekl]{mathbbol}
\usepackage{amsmath,dsfont,mathtools}
\usepackage{amssymb}
\usepackage[normalem]{ulem}
\usepackage{color}
\usepackage{stmaryrd}
\usepackage{adjustbox}
\usepackage{physics}
\usepackage{cleveref}
\usepackage{amsfonts}
\usepackage{soul} 
\usepackage{todonotes}
\usepackage{overpic}
\usepackage{authblk}
\usepackage{pgf}
\usepackage{caption}
\usepackage{subcaption}

\usepackage{amsthm}
\usepackage{comment}
\usepackage{tabularx}
\usepackage{algorithm}
\usepackage[noend]{algpseudocode}

\usepackage{thm-restate}

\usepackage{thmtools}
\usepackage{thm-restate}
\usepackage{hyperref}
\usepackage{xcolor}
\newcommand{\figref}[2][]{\hyperref[#2]{\ref*{#2}#1}}
\newcommand{\R}{\mathbb{R}}

\definecolor{border_green_encoding}{HTML}{3C552D}
\definecolor{green_encoding}{HTML}{B8E986}
\definecolor{border_blue_ansatz}{HTML}{154239}
\definecolor{blue_ansatz}{HTML}{7FBFB2}
\definecolor{blue1}{HTML}{548ea1}
\definecolor{blue2}{HTML}{7f88bf}
\definecolor{blue3}{HTML}{7fbfb2}
\definecolor{grayplot}{HTML}{d6d6d6}

\newtheorem{definition}{Definition}

\definecolor{customblue}{HTML}{17178B}

\title{Subspace Preserving Quantum Convolutional Neural Network Architectures}

\author[1,2]{Léo Monbroussou}
\author[3,4]{Jonas Landman}
\author[1,5]{Letao Wang}
\author[1]{Alex Grilo}
\author[1,3]{Elham Kashefi}

\affil[1]{Laboratoire d’Informatique de Paris 6, CNRS, Sorbonne Université, 4 Place Jussieu, 75005 Paris, France}
\affil[2]{CEMIS, Direction Technique, Naval Group, 83190 Ollioules, France}
\affil[3]{School of Informatics, University of Edinburgh, United Kingdom}
\affil[4]{QC Ware, Palo Alto, USA and Paris, France}
\affil[5]{Ecole Normale Supérieure Paris-Saclay, 4 avenue des Sciences, 91190 Gif-sur-Yvette, France}
\begin{document}

\maketitle

\begin{abstract}
Subspace preserving quantum circuits are a class of quantum algorithms that, relying on some symmetries in the computation, can offer theoretical guarantees for their training. Those algorithms have gained extensive interest as they can offer polynomial speed-up and can be used to mimic classical machine learning algorithms. In this work, we propose a novel convolutional neural network architecture model based on Hamming weight preserving quantum circuits. In particular, we introduce convolutional layers, and measurement based pooling layers that preserve the symmetries of the quantum states while realizing non-linearity using gates that are not subspace preserving. Our proposal offers significant polynomial running time advantages over classical deep-learning architecture. We provide an open source simulation library for Hamming weight preserving quantum circuits that can simulate our techniques more efficiently with GPU-oriented libraries. Using this code, we provide examples of architectures that highlight great performances on complex image classification tasks with a limited number of qubits, and with fewer parameters than classical deep-learning architectures.
\end{abstract}
\section{Introduction}

Quantum Machine Learning (QML) has become a promising area for real world applications of quantum computers, but near-term methods and their scalability are still important research topics. A consequent amount of efforts has been put into understanding how to avoid Barren Plateaus (BP) \cite{McClean2018, larocca2024}, a vanishing gradient phenomenon that prevents the variational algorithms \cite{Cerezo2020} from being trained efficiently. In particular, evidence has recently been shown \cite{cerezo2024does} that the structures that allow us to avoid BP seem to allow classical simulation techniques. In addition, other important questions must be tackled to design near-term quantum algorithms that may offer an advantage. How to ensure that the performance of the algorithms will scale with its dimension, and how to compare classical and quantum algorithms on different figures of merit for a same use case? Those questions are especially hard to answer as we only have access to simulation tools over a low number of qubits or to noisy intermediate scale quantum computers \cite{Preskill_2018}. 

In this context, we propose a new QML algorithm that behaves as a Convolutional Neural Network (CNN) architecture \cite{OShea2015AnIT}. This type of neural network is particularly useful in classical Machine Learning for many use cases including, for example, computer vision tasks \cite{LeCun_LeNet, NIPS2012_c399862d}, and Time Series analysis \cite{Younis2022}. We illustrate ours on image classification. This analogy allows to ensure that our algorithm will still be useful on a larger scale, as CNN is widely used. To design such an algorithm, we used Hamming weight (HW) preserving quantum circuits \cite{monbroussou2023}, a particular type of subspace preserving framework that allows one to avoid BP \cite{Larocca2021, diaz2023showcasing, Ragone2023, Fontana2023, monbroussou2023} by considering a Hilbert space of polynomial size at the cost of having only a polynomial advantage, that could be of high degree. While recent works \cite{cerezo2024does, goh2023liealgebraic} have shown that the absence of BP in this framework leads to the existence of efficient simulation under certain conditions, i.e., no exponential running time complexity advantage, we play within this framework and measurement based techniques to offer significant advantages to our method. In addition, large simulations using GPU clusters show impressive results for our method in comparison with the classical one, including a reduction of the number of parameters which could lead to an even greater advantage.

\paragraph{Related Work:} Our proposal differs from previous CNN architectures such as \cite{Cong_2019, kerenidis2019, wei2021quantumconvolutionalneuralnetwork} in several ways. First, it mimics classical convolutional layers, and pooling by using a specific subspace preserving encoding. Therefore, we believe CNNs could be replaced by our quantum equivalents even for large architecture. Secondly, our proposal offers polynomial speed-ups and is therefore "classically simulable", in the sense that a classical algorithm can perform the same computation in polynomial time due to the use of quantum circuits that are subspace preserving. This choice is motivated by the theoretical guarantees of such circuits in the training and expressivity of the model \cite{monbroussou2023,Fontana2023,Ragone2023}. However, one could use our methods to offer a polynomial speed-up of high degree that could eventually achieve a quantum utility in comparison with CNNs, especially considering that our method seems to achieve similar performances with less parameters. Finally, we tackle the classical simulation of our layers by offering a subspace preserving simulation library \cite{Monbroussou_Hamming_Weight_Preserving_2024} that allows us to test our proposal on larger learning problems than usually presented in the QML community. Our proposed QCNN architecture is quite different from the one in \cite{Cong_2019}, and the results found in \cite{bermejo2024} likely do not directly translate to this work. In particular, our architecture contains correlated parameters and measurement-controlled operations. Future work may determine if LOWESA or similar algorithms based on Pauli propagation can be adapted to achieve speed-ups paralleling those of our specialized method. In addition, recent work \cite{bermejo2024} on classical simulation of one specific type of Quantum Convolutional Neural Networks (QCNN) has shown, using the LOWESA algorithm \cite{rudolph2023,fontana2023classicalsimulationsnoisyvariational}, that the Iris Cong \cite{Cong_2019} proposal of QCNN is effectively classically simulable by considering the subspace of the low-weight measurement operators that are sufficient for the classification of “locally-easy” datasets. This method of classical simulation is different from the one used in our library, and could be applied to algorithms that are not subspace preserving. Our proposed QCNN architecture is quite different from the one in \cite{Cong_2019}, and the results found in \cite{bermejo2024} likely do not directly translate to this work. In particular, our architecture contains correlated parameters and measurement-controlled operations. Future work may determine if LOWESA or similar algorithms based on Pauli propagation can be adapted to achieve speed-ups paralleling those of our specialized method, especially for challenging classification tasks such as described in this paper.

In this paper we mainly use operations that are HW preserving. We recall the main properties of those operations in Appendix \ref{chap:HWPreserving_Resume}. In Section \ref{sec:QCNN_Architecture}, we introduce the different HW preserving layers used to design quantum convolutional neural network architectures. First, we present in \ref{subsec:Conv_layer} the Convolutional layer, then in \ref{subsec:pooling_layer} the measurement based pooling layer, and in \ref{subsec:dense_layer} we present the dense layer used at the end of the architecture. In Section \ref{sec:Results_Simu} we discuss the advantage of our quantum methods over its classical analogs in \ref{subsec:Complexity} by focusing on the model complexity, and we present a training comparison in \ref{subsec:Simulations} for image classification tasks used to benchmark CNN architectures using our simulation tools from \cite{Monbroussou_Hamming_Weight_Preserving_2024}. Finally, we conclude in Section \ref{sec:Conclusion}.

\section{Quantum and Classical Convolutational Neural Network Architecture}\label{sec:QCNN_Architecture}

\begin{figure}[h]
    \centering
    \captionsetup{justification=centering}
    \includegraphics[width=0.85\textwidth]{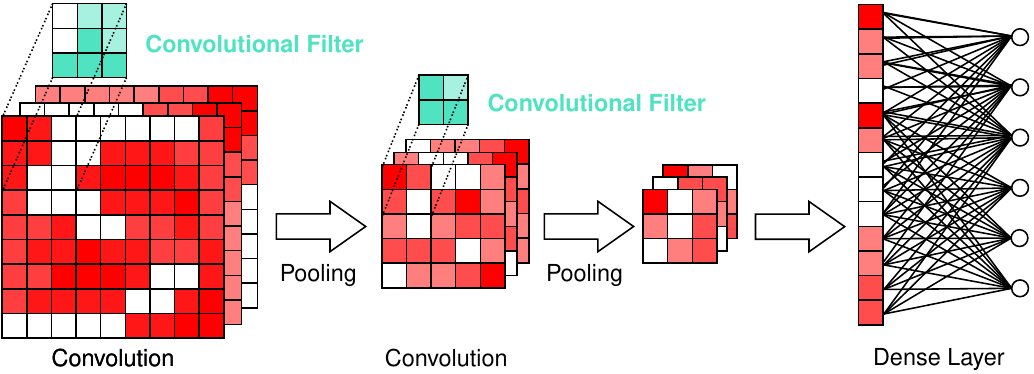}
    \caption{A Convolutional Neural Network architecture. In this example, the input is a batch of 2-dimensional images and is thus a 3-dimensional tensor.}
    \label{fig:figure_intro}
\end{figure}

In this Section, we present our HW preserving convolutional architecture. There exists many type of CNN architectures, and we recall the very first one introduced by LeCun \cite{LeCun_LeNet} which is the original version of LeNet in Figure \ref{fig:figure_intro}. This neural network is composed of successive convolutional and pooling layers, and it ends with a dense layer. 

This structure is quite simple: the convolution parts extract features from the initial images, the pooling parts reduce the dimension of the images, and the final dense layer mixes the features and performs the classification task. The convolution layers are very appropriate for the feature extraction, as they perform a translation invariant operation on the initial image by applying a convolution filter that is optimized through the training. Usually \cite{LeCun_LeNet,NIPS2012_c399862d, he2015deepresiduallearningimage}, each layer can be followed by the application of a nonlinear function.

\subsection{Quantum Convolutional Layer}\label{subsec:Conv_layer}

In this Section we explain our convolutional layer based on tensor encoding. In the following, we will present the classical convolutional layer, and introduce our Hamming weight preserving quantum convolutional. We will show how the quantum version performs a convolution operation that is analog to the classical one, and what their differences are. We illustrate both operations in Figure \ref{fig:Classical_Quantum_Convolutional_Layers}.

\subsubsection*{Classical Convolutional Layer:}

\begin{figure}[h!t]
\centering
\begin{subfigure}[t]{.49\textwidth}
    \centering
    \includegraphics[height=0.6\textwidth]{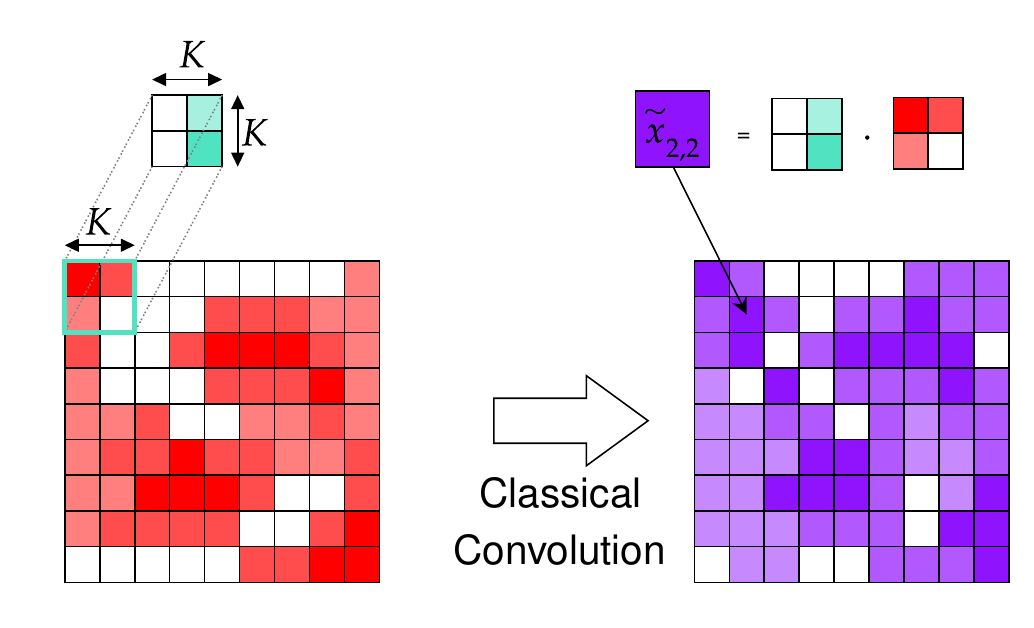}
    \caption{}
    \label{fig:Classical_Conv}
\end{subfigure}
\begin{subfigure}[t]{.49\textwidth}
    \centering
    \includegraphics[height=0.6\textwidth]{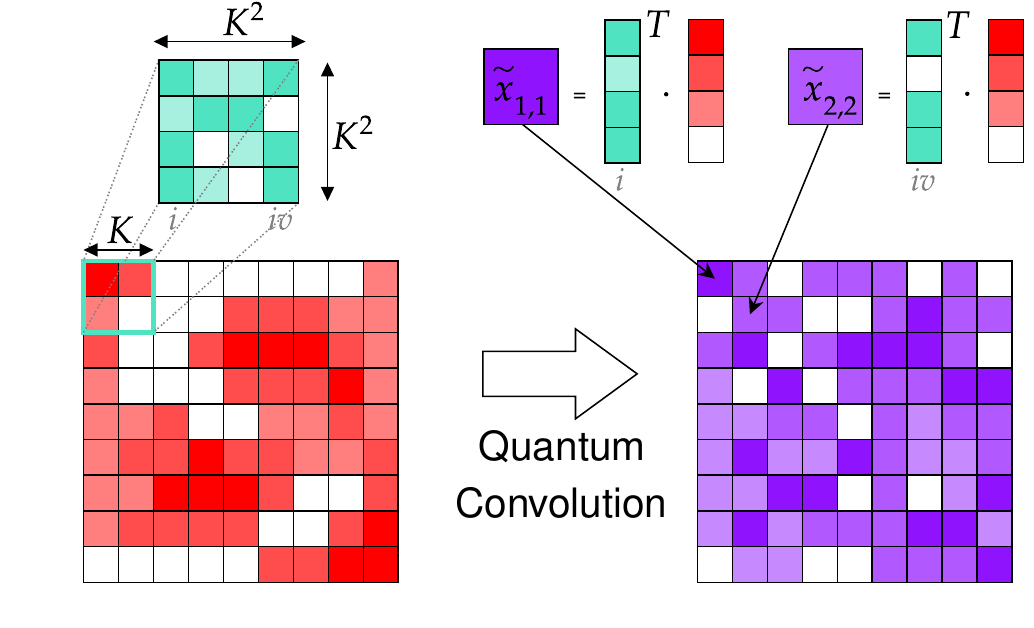}
    \caption{}
    \label{fig:Quantum_Conv}
\end{subfigure}
\caption{Classical (a) and Quantum (b) Convolutional layers. The convolutional filter is represented in blue. }
\label{fig:Classical_Quantum_Convolutional_Layers}
\end{figure}

Let's recall what mathematical operation a classical convolutional layer performs. Consider a 2-dimensional tensor $x = \left(x_{i,j}\right)_{(i,j) \in [d_1]\times[d_2]}$, a convolution filter $W = \left(w_{i,j}\right)_{i,j \in [K]}$ and final image $\Tilde{x} = \left(\Tilde{x}_{i,j}\right)_{(i,j) \in [d_1]\times[d_2]}$. We have:
\begin{equation}\label{eq:classical_2Dconv}
    \forall (i,j) \in [d_1] \times [d_2], \quad \Tilde{x}_{i,j} = \sum_{a, b \in [K]}  w_{a,b} \; x_{i - \lfloor \frac{K}{2} \rfloor + a,  j - \lfloor \frac{K}{2} \rfloor + b} 
\end{equation}
which corresponds to a convolution operation between the Filter tensor and the Filter window around the pixel. In Figure \ref{fig:Classical_Conv}, we illustrate this 2-dimensional example with the Filter window in green and the Filter in blue. We can extend this definition to any $k$-dimensional tensor and for any convolutional layer of dimension less or equal to $k$. Notice that in the case of a 2-dimensional convolution for a 3-dimensional input such as a batch of square images (see Figure \ref{fig:figure_intro}), each image is affected by the same 2-dimensional convolutional operation with the filter, such as described in Eq.\eqref{eq:classical_2Dconv}.

    \subsubsection*{Tensor Encoding}

To perform the quantum convolutional layer and the encoding, we will use Reconfigurable Beam Splitter (RBS) gates, well-known two qubit gates used for HW preserving algorithms \cite{Landman2022, Cherrat2022, jain2023quantumfouriernetworkssolving}. Additional information and properties of this gate can be found in Appendix \ref{chap:HWPreserving_Resume}.

\begin{definition}[Reconfigurable Beam Splitter gate]\label{def:RBSdef}
The Reconfigurable Beam Splitter (RBS) gate is a 2-qubit gate that corresponds to a $\theta$-planar rotation between the states $\ket{01}$ and $\ket{10}$:
    \begin{equation}\label{eq:RBS_2_qubit_gate}
    RBS(\theta) = \begin{pmatrix}
        1 & 0 & 0 & 0 \\
        0 & \cos(\theta) & \sin(\theta) & 0 \\
        0 & -\sin(\theta) & \cos(\theta) & 0 \\
        0 & 0 & 0 & 1 \\
        \end{pmatrix} \, \textrm{.}
    \end{equation}    
\end{definition}

We propose encoding classical data in such a way that allows us to apply a convolutional layer by using HW preserving circuits. More precisely, we propose to load any tensor of dimension $k$ by using amplitude encoding on a Tensor Basis of HW $k$. 

\begin{definition}[HW Preserving Tensor encoding]\label{def:TensorEncoding}
    Consider a classical tensor of dimension $k$ such that $x = (x_{1, \dots, 1}, \dots,  x_{d_1, \dots, d_k}) \in \R^{d_1 \times \dots \times d_k}$. An amplitude encoding tensor encoding data loader is a parametrized $n$-qubit quantum circuit (with $n = \sum_{i \in [k]} d_i$ that prepares the quantum states:
    \begin{equation}
        \ket{x} = \frac{1}{||x||} \sum_{i_1 \in [d_1]} \dots \sum_{i_k \in [d_k]}  x_{i_1, \dots, i_k} \ket{e^{d_1}_{i_1}} \otimes \dots \otimes \ket{e^{d_k}_{i_k}} ,
    \end{equation}
    where $\ket{e_{i_l}^{d_l}} = \ket{0 \dots 0 1 0 \dots 0}$ is a state corresponding to a bit-string with $d_l$ bits and only the bit $i_l$ is equal to $1$. Therefore, for any $j \in [k]$, $\left\{\ket{e_{i}^{d_l}} \mid i \in [d_l] \right\}$ is a fixed family of $d$ orthonormal quantum states, and $||\cdot||$ denotes the $2$-norm of $\mathbb{R}^d$.
\end{definition}

For example we can map a $2 \times 2$ matrix image $x$ to a state $\ket{x}$ using this encoding:
\begin{equation}
    X = \begin{pmatrix} x_{1,1} & x_{1,2} \\
                        x_{2,1} & x_{2,2} 
        \end{pmatrix} \; \longrightarrow \; \ket{x} = \frac{1}{||x||} \left( x_{1,1}\ket{1010} + x_{1,2}\ket{1001} + x_{2,1}\ket{0110} + x_{2,2}\ket{0101}\right) \, \textrm{.}
\end{equation}

This choice of this encoding gives a structure to the state that allows us to apply the Convolutional layer and the Pooling layers described in the following. It can be considered as amplitude encoding on a specific basis, and can be realized thanks to quantum data loaders that perform amplitude encoding on the basis of fixed HW \cite{farias2024, Johri2020, monbroussou2023}. The tensor encoding offers the opportunity to use measurement based operation to apply a Pooling operation as described in \ref{subsec:pooling_layer} that reduces the dimension of the state and apply non-linearities while preserving the tensor encoding structure of the final states which has never been done before to our knowledge. This is the key ingredient for our global convolutional architecture. Notice that the final state can be decomposed on $k$ registers, and that all the registers can be considered alone as HW preserving circuits of HW $1$.

\subsubsection*{Hamming-Weight Preserving Convolutional Layer:}

Considering a tensor encoding of dimension $k$, applying a RBS-based quantum circuit on $K$ qubits of one register performs rotations between the states corresponding to each pixel linked with those qubits. As an example, performing a RBS-based quantum circuit on the $K$ first qubits of the line register for a 3 dimension image tensor encoded will affect all the pixels in the $K$ first lines of all the images. By applying the same circuit to each $K$ consecutive qubits of each register, one can perform a $k$-dimensional convolution. On each register, the HW is equal to $1$ (or unary). 

For example, with $k=2$, we consider a 2-dimensional tensor $x = \left(x_{i,j}\right)_{(i,j) \in [d_1]\times[d_2]}$ which is tensor encoded such as in Definition \ref{def:TensorEncoding}. If one applies a RBS based circuit between all the qubits of indexes $I, \dots, (I+K) \in [d_1]$ of the line register, and another one between all the qubits of indexes $J, \dots, (J+K) \in [d_2]$ of the column register, then one can consider that the corresponding $K \times K$ pixels form a filter window affected by a unitary matrix $U_\textrm{Filter}(\Theta)$ such that:
\begin{equation}\label{eq:quantum_conv_unitary}
    \sum_{i=I}^{I+K} \sum_{j=J}^{J+K} \Tilde{x}_{i,j} \ket{e_i, e_j} = U_{\textrm{Filter}}(\Theta) \sum_{i=I}^{I+K} \sum_{j=J}^{J+K} x_{i,j} \ket{e_i, e_j} \, \textrm{,}
\end{equation}
with $\Theta$ the RBS parameters, $\ket{e_i}$ is a unary state corresponding to the Definition \ref{def:TensorEncoding}, $U_{\textrm{Filter}}(\Theta) = (u_{i,j}(\Theta))_{i,j \in [K^2]}$ the quantum convolutional filter, and the final image $\Tilde{x} = \left(\Tilde{x}_{i,j}\right)_{(i,j) \in [d_1]\times[d_2]}$ which is still tensor encoded.

Each pixel in the convolutional window is affected by the quantum filter by a convolutional relation analog to the one given by Eq.\eqref{eq:classical_2Dconv}:
\begin{equation}\label{eq:quantum_conv}
    \forall i \in \llbracket I, I+K \rrbracket, \; \forall j \in \llbracket J, J+K \rrbracket, \quad \Tilde{x}_{i,j} = \sum_{a,b \in [K]}  w^{i,j}_{a,b} \; x_{a,b} \quad \textrm{with} \; w^{i,j}_{a,b} = \bra{e_a, e_b} U_{\textrm{Filter}}(\Theta) \ket{e_i,e_j} \, \textrm{.}
\end{equation}
We show through Eq.\eqref{eq:quantum_conv}, that applying a same RBS-based circuit to each K consecutive qubits for each register of a tensor encoded state is equivalent to applying a convolution function. The quantum convolution is analog to the classical one in the sense where each pixel in the filter window is affected by a classical convolution operation with a $K \times K$ classical filter corresponding to a part of the $K^2 \times K^2$ quantum filter coefficients. The HW preserving convolutional layer is illustrated in Figure \ref{fig:Quantum_Conv}. Notice that we apply the same operations to each set of $K \times K$ pixels, and not the same operation to each pixel. This limitation can be bypassed by loading several copies of the initial tensor that are translated in a batch, we implement such a feature in \cite{Monbroussou_Hamming_Weight_Preserving_2024}. Eq.\eqref{eq:quantum_conv_unitary} and Eq.\eqref{eq:quantum_conv} can be adapted for any initial tensor dimension and any filter dimension.

For a $d$-dimensional quantum convolutional layer, the quantum filter unitary is of size $K^d \times K^d$. We use $K$ as the quantum convolutional layer hyperparameter. Even if the quantum filter is bigger than the classical filter, the structure of the quantum convolutional circuit is such that the number of parameters is smaller for the quantum layer, as explained in Section \ref{subsec:Complexity}. We show in our simulations presented in Section \ref{subsec:Simulations} that the QCNN architecture performances are similar to the ones of classical CNN architecture.

\begin{figure}[h]
    \centering
    \includegraphics[width=0.95\textwidth]{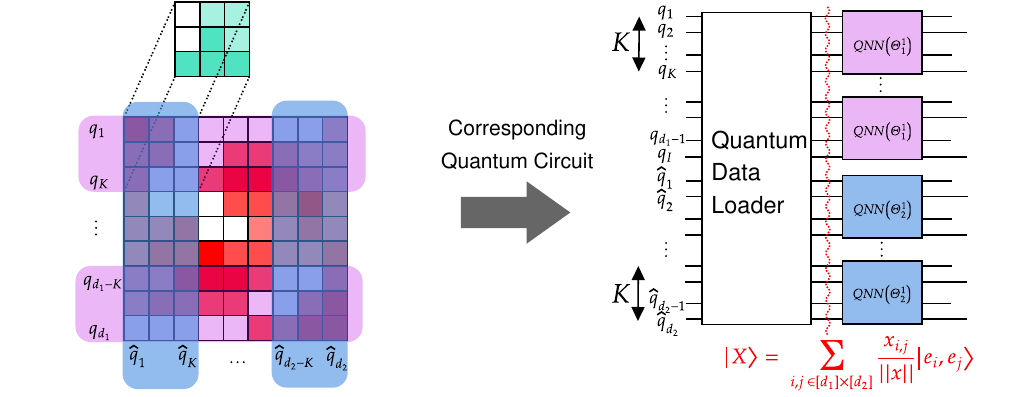}
    \caption{A 2 dimensional convolutional layer using HW preserving quantum circuits and tensor encoding.}
    \label{fig:Convolutional_Layer}
\end{figure}

In this setting, previous work \cite{Landman2022, Cherrat2022} have proposed efficient circuits that maximize the controlability with different depth, meaning that those circuits can reach ion the unary basis any orthogonal matrices. In the case of the tensor encoded data, each register has a corresponding HW of $1$. Usual HW preserving ansatz that are can achieve any unitary matrix in shte subspace of HW $1$ are presented in Appendix \ref{chap:HWPreserving_Ansatz}. In the following, we will focus on the butterfly circuit \cite{Cherrat2022} that minimizes the depth.  Therefore, considering a $d$-dimensional quantum filter of size $\prod_{i=1}^d K_i$, the depth of one convolutional layer is $ \max_{i \in [d]} \mathcal{O}\left( \log(K_i) \right)$. We illustrate the HW preserving convolutional layer in Figure \ref{fig:Convolutional_Layer}. Notice that the $d$-dimensional filter that we apply in the convolutional layer cannot be any matrix of size $K_1 \dots K_d$. As explained in \cite{Landman2022}, $n$-qubit RBS based circuits can only perform $n \times n$ orthogonal transformations while considering a HW of $1$. The resulting filter, or equivalent unitary on all the registers, is thus a parametrized orthogonal matrix with $\sum_{i=1}^d \frac{K_i (K_i -1)}{2}$ independent parameters.

In terms of time complexity, the classical CNN layer depends on the input size, and its filter size. While considering a batch of $I \times I$ images, the complexity of conventional 2D convolution \cite{Wei2021RethinkingCT} depends on the size of the input image, the number of channels $C$, and is $\mathcal{O}\left( C^2 \cdot K^2 \cdot I^2\right)$ (we consider that the final are of dimension $I \times I$). For our HW preserving convolutional layer, the complexity only depends on the filter size $K$. As explained previously, considering a butterfly circuit that maximized the expressivity in the subspace of HW $1$, the depth of the quantum circuit is $\mathcal{O}\left(\log(K)\right)$. A comparison of the number of parameters and the time complexity between classical and quantum convolutional architecture layers is presented in Section \ref{subsec:Complexity}. The quantum polynomial advantage increases with the dimension of the tensor, i.e., the HW of the encoding. Therefore, this layer may offer a more interesting advantage for use cases with inputs of large dimensions, such as series classification \cite{Ismail_Fawaz_2019}.
\subsection{Quantum Pooling Layer}\label{subsec:pooling_layer}

In this Section, we introduce a Pooling layer that preserves the tensor structure of the quantum state. This layer allows us to reduce the dimension of the image but also to apply some non linearities by using measurements. Applying non linearities in QML architectures is a non trivial task as variational quantum circuits perform linear algebraic operations on the quantum state. Previous works propose to use classical computation between quantum layers to apply non linearities \cite{Landman2022} or to use specific hardware tools to create non-linearities \cite{steinbrecher2018quantum}. Another proposal \cite{Cong_2019} of quantum CNN has considered using measurements and single qubit gates controlled by the outcomes to perform non linearities. However, to our knowledge, there is no existing method to perform a Pooling layer with non linearities that preserves the structure of the state, allowing to keep the subspace preserving properties of the computation. Therefore, our proposal offers the possibility of deep learning architectures that are subspace preserving and thus, that ensure theoretical guarantees on their training. In addition, our method does not require using adaptive measurement techniques, but only to consider CNOT gates and to ignore a part of the qubits in the remaining part of the circuit.

\begin{figure}[h]
    \centering
    \includegraphics[width=0.95\textwidth]{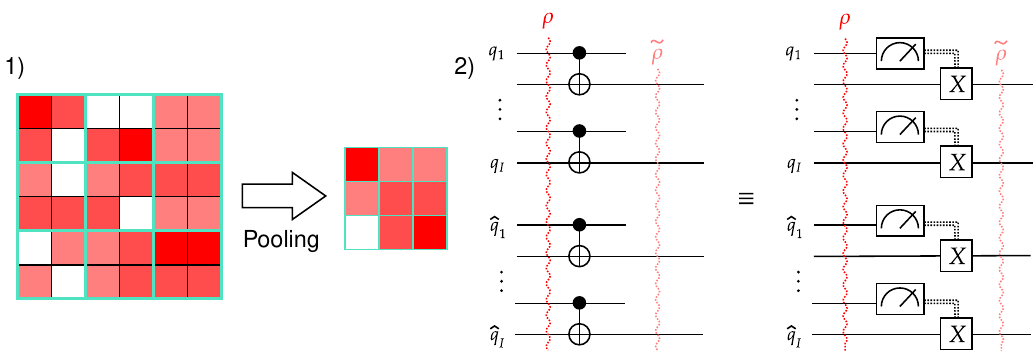}
    \caption{1) Illustration of the Pooling Layer effect on a 2 dimensional image with Pooling windows in blue. 2) Quantum Circuit for the Pooling layer and its equivalent representation using measurement and control X gates.}
    \label{fig:Pooling_layer}
\end{figure}

Considering the tensor encoding, our Pooling method consists in applying a CNOT between each pair of qubits in the register corresponding to the dimension that we want to reduce. In the following part of the circuit, we will only consider the target qubits. This method is mathematically equivalent to measuring the control qubits and applying a bit flip operation to the target qubits when measuring the corresponding control qubits in state $\ket{1}$.

This Pooling circuit preserves the tensor encoding structure. Considering an initial state $\rho = \ket{X}\bra{X}$ with $\ket{X}=\sum_{i,j \in [I]}\frac{x_{i,j}}{||x||} \ket{e^I_i} \bigotimes \ket{e^I_j}$, the resulting state considering a Pooling layer for a square image (see Figure \ref{fig:Pooling_layer}) is:
\begin{equation}
    \Tilde{\rho} = \sum_{i} p_i \ket{\Tilde{X}^i} \bra{\Tilde{X}^i} \quad \text{with} \quad \ket{\Tilde{X}^i} = \sum_{l,k \in [O]}\frac{\Tilde{x}^i_{l,k}}{||x||} \ket{e^O_l} \bigotimes \ket{e^O_k} \, \textrm{,}
\end{equation}

with $O = I/2$. The preservation of the tensor encoding structure allows us to implement several convolutional and pooling layers as in most classical deep learning architecture. In addition, this pooling operation is analog to the average pooling operation commonly used in deep learning architecture. Consider the case of a 4 by 4 image in which we apply this pooling operation:
\begin{equation}
    X = \begin{pmatrix} x_{1,1} & x_{1,2} & x_{1,3} & x_{1,4} \\
                        x_{2,1} & x_{2,2} & x_{2,3} & x_{2,4} \\
                        x_{3,1} & x_{3,2} & x_{3,3} & x_{3,4} \\
                        x_{4,1} & x_{4,2} & x_{4,3} & x_{4,4}
        \end{pmatrix} \rightarrow \Tilde{\rho} \, \textrm{.}
\end{equation}
And
\begin{equation}
    \Tilde{\rho} =
        \begin{pmatrix} x^2_{11} + x^2_{12} + x^2_{13} + x^2_{14} & x_{12} x_{14} + x_{22} x_{24} & x_{21} x_{41} + x_{22} x_{42} & x_{22} x_{44} \\
        x_{12} x_{14} + x_{22} x_{24} & x^2_{13} + x^2_{14} + x^2_{23} + x^2_{24} & x_{24} x_{42} & x_{23} x_{43} + x_{24} x_{44} \\
        x_{21} x_{41} + x_{22} x_{42} & x_{24} x_{42} & x^2_{31} + x^2_{32} + x^2_{41} + x^2_{42} & x_{32} x_{34} + x_{42} x_{44} \\
        x_{22} x_{44} & x_{23} x_{43} + x_{24} x_{44} & x_{32} x_{34} + x_{42} x_{44} & x^2_{33} + x^2_{34} + x^2_{43} + x^2_{44}
        \end{pmatrix} \, \textrm{.}
\end{equation}
Notice that the diagonal terms are the sum of the squared values of the pixels in the Pooling windows (see Figure \ref{fig:Pooling_layer}). Therefore, the probability of measuring the state corresponding to a certain pixel after the Pooling layer is the sum of the probability of measuring the states carrying the pixels in the corresponding Pooling window. Finally, considering a measurement based Pooling operation allows us to apply some non linearities to the quantum state which is good as non-linear activation functions are usually used after the Pooling layers. 
\subsection{Quantum Dense Layer}\label{subsec:dense_layer}

In this Section, we discuss the final part of our subspace preserving deep-learning architecture. As explained in the introduction of Section \ref{sec:QCNN_Architecture}, we consider an architecture very similar to LeNet (see Figure \ref{fig:figure_intro}). After applying several Convolutional and Pooling Layers, this architecture ends with a vectorization of the image followed by a fully connected layer or dense layer.

\begin{figure}[h]
    \centering
    \includegraphics[width=0.95\textwidth]{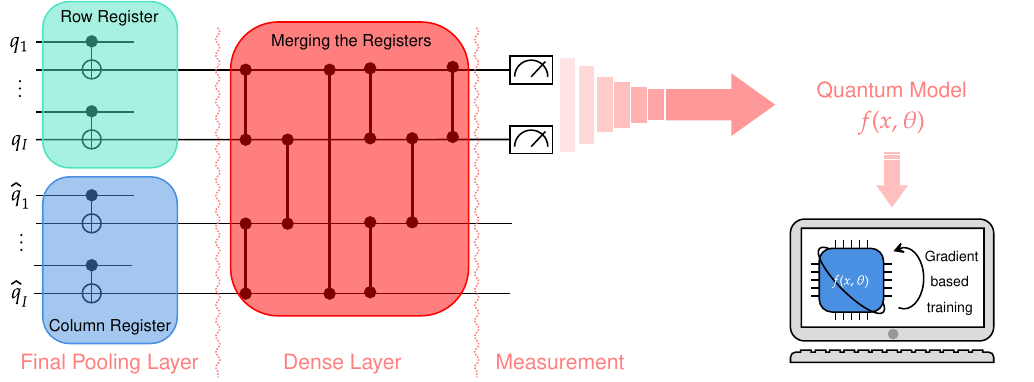}
    \caption{A 2 dimensional convolutional layer using Hamming weight preserving quantum circuits and tensor encoding. Vertical lines in the dense layer represent two-qubit RBS gates, parametrized with independent angles.}
    \label{fig:Dense_Layer}
\end{figure}

In the case of our architecture, no vectorization is required. The dense layer only consists of applying a RBS-based trainable quantum circuit to the remaining qubit while merging all the registers.

Applying such a circuit on a quantum state of fixed HW $k$ corresponds to applying an orthogonal neural network as a dense layer. Previous works \cite{Landman2022, monbroussou2023} have highlighted the fact that using quantum orthogonal neural networks results in powerful neural networks. The number of parameters and the choice of the structure, and more specifically the choice of the connectivity used for this circuit are very important in the maximal controlability of this layer. For example, using a line connectivity for this $n$-qubit layer implies that the layer equivalent unitary is a compound matrix \cite{kerenidis2022qmlsubspace, monbroussou2023}. This reduced the maximal dimension of the Dynamical Lie Algebra of this layer to $n(n-1)/2$, meaning that only a low number of parameters can be useful.

According to the HW of the states in the dense layer, i.e., the dimension $k$ of the tensor considered as the input of the architecture, the dense layer is harder to simulate classically. The impact of a RBS gate parametrized by $\theta$ in a subspace of $n$ qubits and HW $k$ is $\binom{n-2}{k-1}$ $\theta$-planar rotations. Therefore, considering a HW $k$ independent of the number of qubits $n$, the dense layer can be classically simulated. However, the complexity of this simulation could\footnote{to our knowledge, performing the $\binom{n-2}{k-1}$ $\theta$-planar rotations is the best simulation algorithm that exists for RBS-based quantum circuits.} be polynomial of degree $k$.
\section{Results and Simulation}\label{sec:Results_Simu}

    \subsection{Complexity results}\label{subsec:Complexity}

The term "model complexity" may refer to different meanings in deep learning, including the expressive capacity and effective model complexity \cite{hu2021modelcomplexitydeeplearning}. It may also refer to the time complexity of the different layers \cite{Shah2022TimeCI}. To compare the feed-forward and training running time between the quantum deep learning layers introduced in Section \ref{sec:QCNN_Architecture} and their classical equivalent, two important criteria are particularly significant. First, the number of parameters of the model is a standard metric of the running time, as a low number of parameters reduces the training and the forward pass of a model. In addition, the forward pass running time is very important, and determines the number of basic operations a computer needs to run the model. In Table \ref{table:Time_Complexity}, we compare the running time complexity of the forward pass for each convolutional neural network layer with the depth of the analog quantum layers. The depth of the corresponding quantum circuits gives the number of basic quantum operations, i.e., number of parallel gates that should apply.

\begin{table}[h!]
\centering
\begin{tabular}{ |c|c|c|c|c| } 
    \hline
    & Convolutional & Pooling & Orthogonal Dense & Dense \\ 
    \hline
    Classical Time Complexities & $\mathcal{O}( \prod_{i=1}^k K_i^2 \cdot d_i^2)$ & $\mathcal{O}(\prod_{i=1}^k d_i)$  & $\mathcal{O}(p \cdot \binom{n}{k})$ & $\mathcal{O}(\sum_{i=1}^k d_i^2)$ \\ 
    \hline
    Quantum Layer Depth & $\mathcal{O}(\log(K))$ & $\mathcal{O}(1)$ & $\mathcal{O}(\frac{p}{n})$ & - \\
    \hline
\end{tabular}

\caption{Time complexity comparison between classical deep learning layers and Hamming weight preserving quantum analogs. We consider $k$ dimensional convolutional neural network layers with $d_1 \times \cdots \times d_k$ the size of the square input tensor, $\{ K_1, \cdots K_k \}$ the size of the convolutional filter, and $p$ the number of parameters in the orthogonal dense layer. We call $n$ the global number of qubits in the case of the quantum architecture where $n = \sum_{i=1}^k d_i$.}
\label{table:Time_Complexity}
\end{table}

In Section \ref{sec:QCNN_Architecture}, we presented our layers in the case of $2$ or $3$-dimensional convolutional architecture. In Table \ref{table:Time_Complexity}, we consider the case of $k$-dimensional convolutional layer to consider a general case. The quantum advantage increases with the dimension of the tensor. However, this dimension corresponds to the global HW, and one should be careful to consider this value independent of the number of qubits $n$ to avoid Barren Plateau \cite{Larocca2021, diaz2023showcasing, monbroussou2023}. 

In addition to the running time complexity, one should also consider the number of parameters of the model and the running time associated with the vectorizations in the model. Indeed, preparing the state for each layer in a classical CNN architecture requires to vectorized it, especially when using GPUs for computation \cite{ren2015vectorizationdeepconvolutionalneural}. In the case of our quantum models, we don't need to adapt the state as our Convolutional and Pooling layers preserve the structure of our state, and the final dense layer only requires to apply RBS between qubits from different registers. 

\begin{table}[h!]
\centering
%\captionsetup{justification=centering}
\begin{tabular}{ |c|c|c|c|c| } 
    \hline
    & Convolutional & Pooling & Orthogonal Dense & Dense \\ 
    \hline
    Classical Layers & $ \prod_{i=1}^k K_i^2$ & $0$  & $p \leq \binom{n}{k}(\binom{n}{k}-1)/2$  & $(\sum_{i=1}^k d_i)^2$ \\ 
    \hline
    Quantum Layers & $\sum_{i=1}^k K_i(K_i - 1)/2$ & $0$ & $p \leq \binom{n}{k}(\binom{n}{k}-1)/2$ & -  \\
    \hline
\end{tabular}
\caption{The number of parameters of classical deep learning layers and of Hamming weight preserving quantum analogs. We consider $k$ dimensional convolutional neural network layers with $d_1 \times \cdots \times d_k$ the size of the square input tensor, $\{ K_1, \cdots K_k \}$ the size of the convolutional filter. We call $n$ the global number of qubits in the case of the quantum architecture where $n = \sum_{i=1}^k d_i$.}
\label{table:Number_Parameters}
\end{table}

Thanks to Table \ref{table:Time_Complexity} and Table \ref{table:Number_Parameters}, we observe that the Convolutional and Pooling layers offer large polynomial advantages, especially when considering high dimensional input tensors. The quantum filter is less parametrized than the classical one, but simulations presented in Section \ref{subsec:Simulations} show that those quantum orthogonal filters perform well. Similarly, the quantum orthogonal dense layer performs well with a reduced number of parameters in comparison with classical dense. Previous works \cite{Landman2022, Cherrat2022, Johri2020} have already shown that orthogonal layers perform well in comparison with dense layers. \textbf{The quantum advantage in terms of running time complexity, number of parameters, and lack of vectorization needed open new perspective to design useful subspace preserving QML algorithms.}

    \subsection{Simulations}\label{subsec:Simulations}

In this Section, we test our method using several very famous datasets used to benchmark classification algorithms. We proposed in our \cite{Monbroussou_Hamming_Weight_Preserving_2024} a GPU-based toolkit to simulate Hamming-Weight preserving deep-learning architectures. To do so, the code performs linear algebra using the PyTorch \cite{paszke2019pytorchimperativestylehighperformance} library while only considering the smaller subspace possible. The pooling part of the circuit is simulated using projectors between different subspace bases. Thanks to our method, we were able to simulate larger quantum circuits and to perform image classification on $10$-classes datasets and not only binary classification as usually done in QML. Our simulation software allows one to mix our subspace preserving simulation with classical layers thanks to its PyTorch module implementation. To our knowledge, this is the most complex image classification task, in the sense of the number of labels, realized with classical data.

\begin{table}[h!]
\centering
\begin{tabular}{ |c|c|c|c|c|c| } 
    \hline
    & Parameters & Dataset & Training Accuracy &  Testing Accuracy & Epochs \\ 
    \hline
    \multirow{3}*{CNN Architecture} & \multirow{3}*{990}  & MNIST & $91.33\% \pm 0.36\%$ & $84.59\% \pm 0.91\%$ & 30 \\ 
    & & FashionMNIST & $82.8\% \pm 0.3\%$ & $73.83\% \pm 1.56\%$ & 40 \\ 
    & & CIFAR-10 & $35.65\% \pm 0.43\%$ & $27.79\% \pm 0.85\%$ & 40  \\
    \hline
    \multirow{3}*{QCNN Architecture} & \multirow{3}*{755}  & MNIST & $93.79\% \pm 0.76\%$ & $86.79\% \pm 1.45\%$ & 30 \\ 
    & & FashionMNIST & $82.95\% \pm 0.47\%$ & $78.29\% \pm 0.83\%$ & 40\\ 
    & & CIFAR-10 & $34.29\% \pm 1.15\%$ & $28.71 \% \pm 1.05\%$ & 40 \\
    \hline
\end{tabular}
\caption{Simulation Results. We consider $2000$ training samples and $1000$ testing samples. We trained the architectures described in Figure \ref{fig:Simulations}, with Adam optimizer, and Cross Entropy Loss. All hyper-parameters and computations can be found in \cite{Monbroussou_Hamming_Weight_Preserving_2024}.}
\label{table:Simulations}
\end{table}

\begin{figure}[h!t]
\centering
\begin{subfigure}[t]{.45\textwidth}
    \centering
    \includegraphics[height=0.65\textwidth]{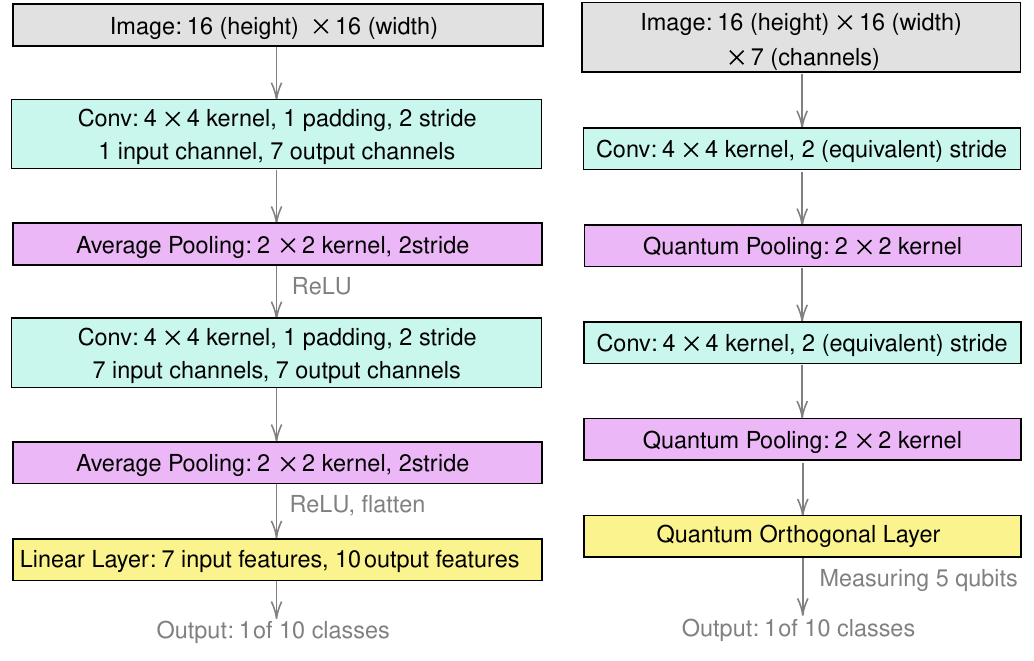}
    \caption{CNN architecture and HW preserving convolutional architecture used for training comparison.}
    \hspace*{0.2in}
    \label{fig:QCNN_CNN_Architecture}
\end{subfigure}
\begin{subfigure}[t]{.45\textwidth}
    \centering
    \includegraphics[height=0.75\textwidth]{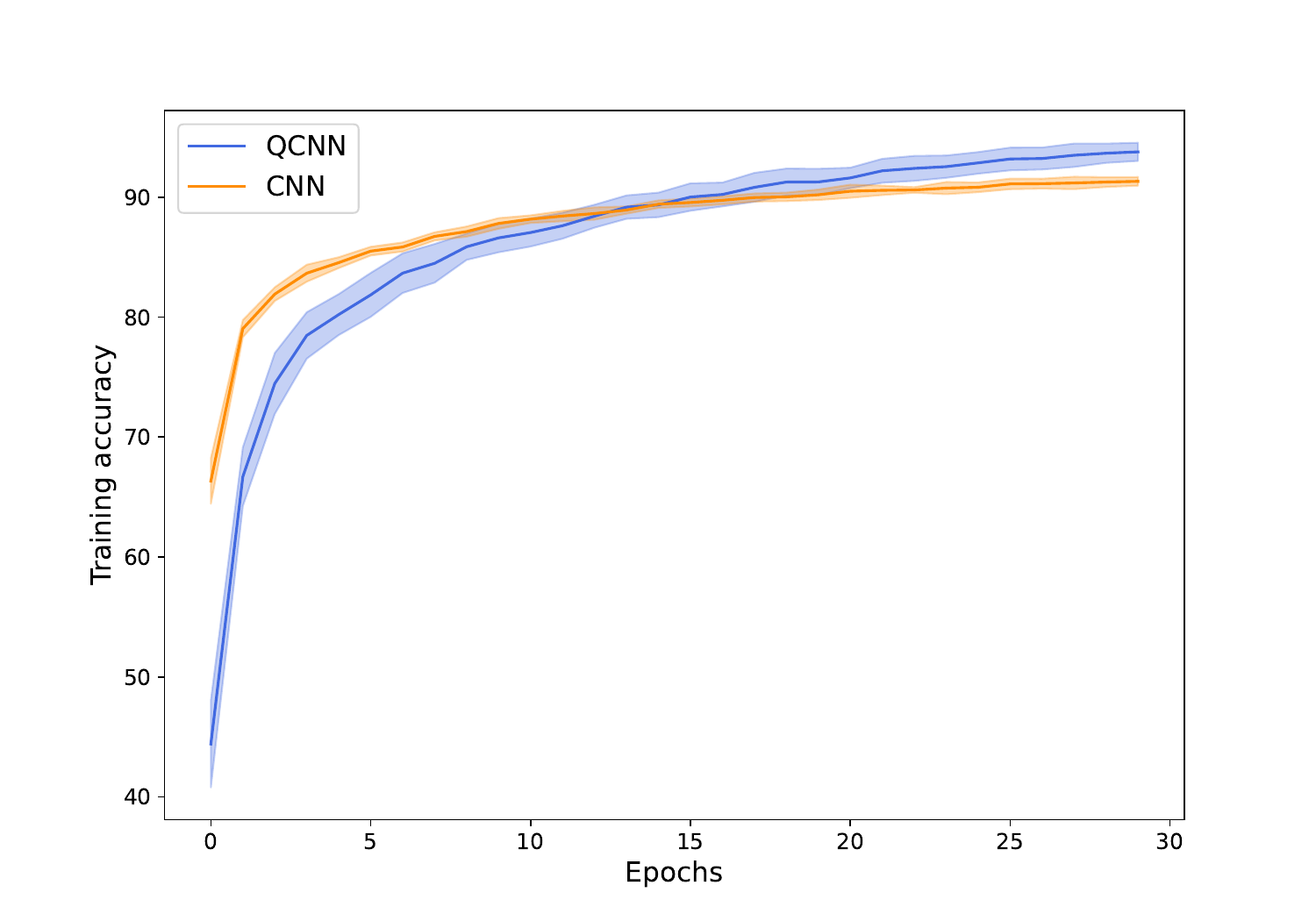}
    \caption{MNIST digit dataset \cite{lecun2010mnist}. 
    }
    \label{fig:QCNN_CNN_MNIST}
\end{subfigure}
\begin{subfigure}[t]{.49\textwidth}
    \centering
    \includegraphics[height=0.75\textwidth]{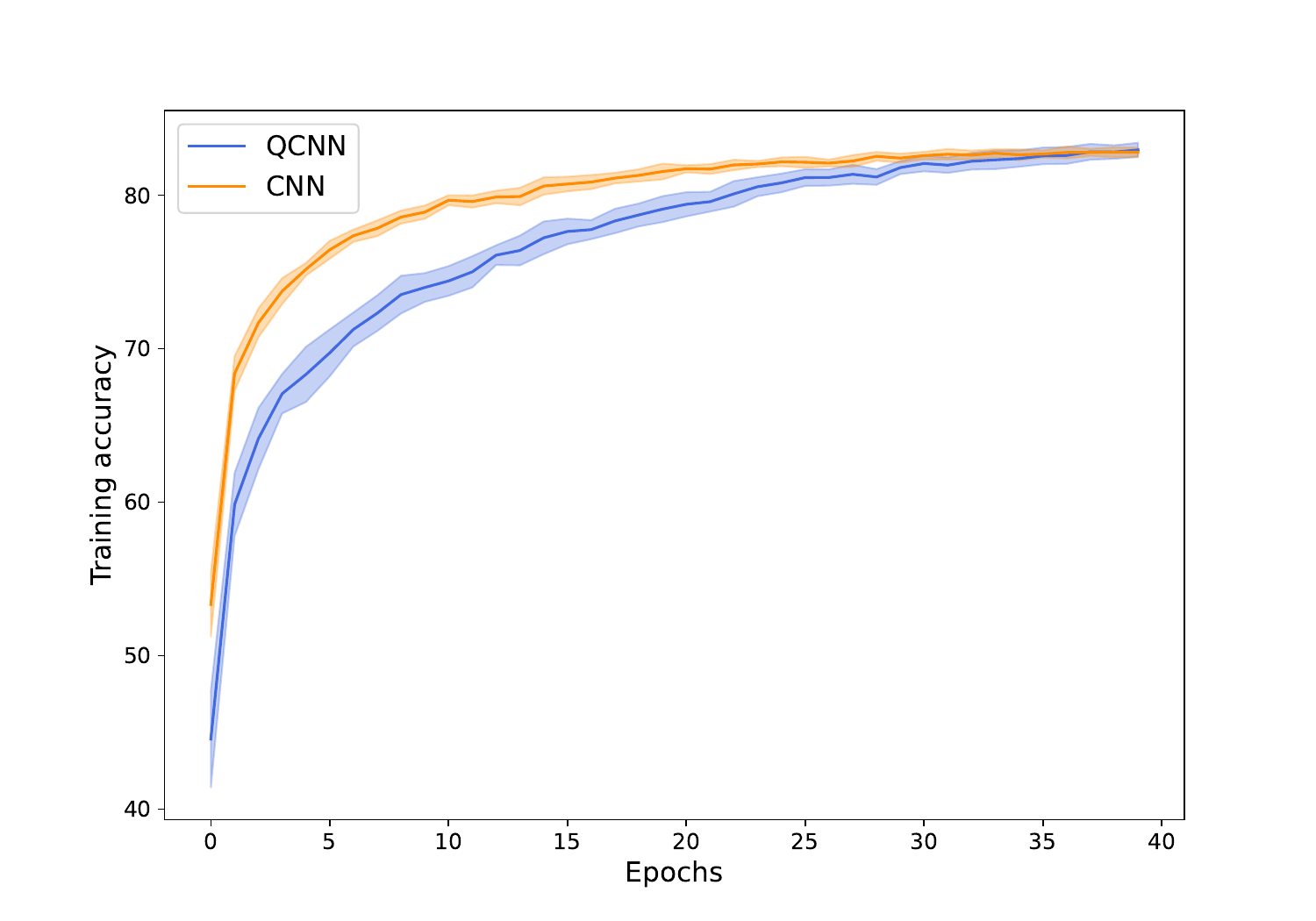}
    \caption{Fashion MNIST dataset \cite{xiao2017fashionmnistnovelimagedataset}. }
    \hspace*{.2in}
    \label{fig:QCNN_CNN_FashionMNIST}
\end{subfigure}
\begin{subfigure}[t]{.49\textwidth}
    \centering
    \includegraphics[height=0.75\textwidth]{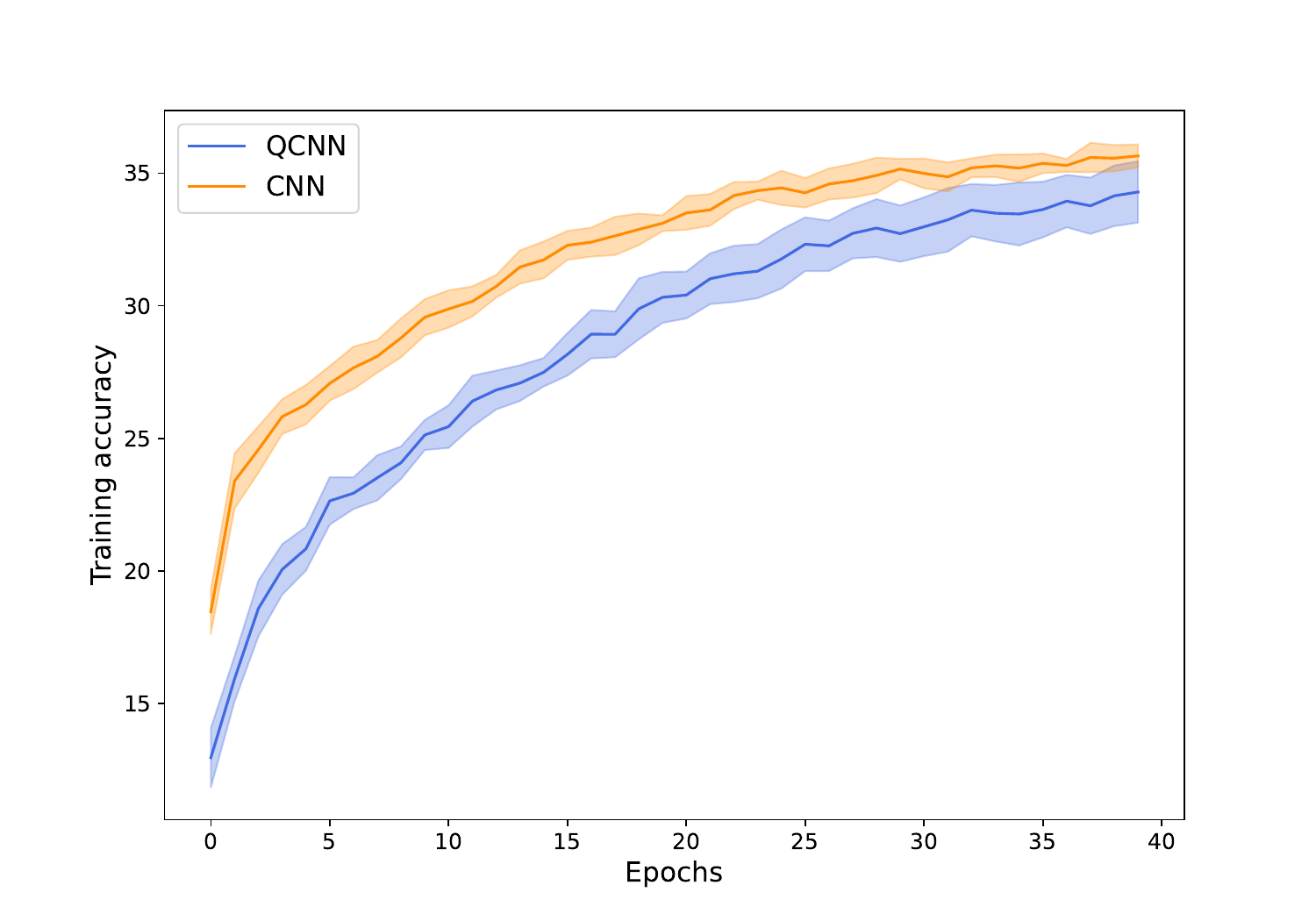}
    \caption{CIFAR-$10$ dataset \cite{Krizhevsky2009LearningML}.}
    \vfill
    \label{fig:QCNN_CNN_CIFAR}
\end{subfigure}
\caption{Average training accuracy and standard deviation comparison between classical CNN architecture and a HW preserving architecture (a) for classification of 10 label datasets (b,c,d), with $2000$ input images. The average values and standard deviation are derived from 10 different trainings. The quantum architecture (QCNN) has $755$ parameters and the classical architecture (CNN) has $990$ parameters. Both architecture (choice of layers and hyper-parameters) are unchanged in all the case, and can be found in \cite{Monbroussou_Hamming_Weight_Preserving_2024}.}
\label{fig:Simulations}
\end{figure}

To benchmark our layers, we propose to compare a classical CNN architecture with a quantum one and similar hyper-parameters, for $4$ well known image recognition datasets. Each dataset (\cite{lecun2010mnist,xiao2017fashionmnistnovelimagedataset,Krizhevsky2009LearningML} has $10$ classes of image, which we prepare by applying a average pooling layer to reduce the size of the input images. Every simulation can be found in \cite{Monbroussou_Hamming_Weight_Preserving_2024}, and an illustration of both architectures is presented on Figure \ref{fig:QCNN_CNN_Architecture}. We ran our simulations using a NVIDIA A100 80 GB GPU on a cluster.

Results presented in Figure \ref{fig:Simulations} and Table \ref{table:Simulations} show that our architecture offers similar performance than classical CNN architecture. In addition, with the running time complexity advantages summarized in Section \ref{subsec:Complexity}, and the lack of vectorization needed, the quantum architecture reaches similar accuracy with fewer parameters due to the orthogonality of its final dense layer, and the structure of its convolutional layers. Our model even outperforms the classical architecture for the MNIST and Fashion MNIST dataset classification. In the case of CIFAR-10 dataset, both architectures do not have the complexity to achieve a satisfying result after the training, but we observe similar training behavior and performance. 

\newpage
 
\section{Conclusion}\label{sec:Conclusion}

In this paper, we introduce a Convolutional and a measurement based Pooling layer that offer polynomial advantages over their classical analogs. By conserving the subspace preserving structure of the state during the computation, these layers can be assembled to perform complex deep-learning algorithms such as Convolutional Neural Network architectures, while assuring the correct training of the quantum circuit. In particular, those circuits can avoid Barren Plateau by only considering subspaces of polynomial size, limiting the potential running time advantages to polynomial ones. Recent work \cite{cerezo2024does} has pointed out the link between the absence of Barren Plateau and a non-exponential advantage in the near-term QML literature, and we believe that our proposal offers a promising path for useful QML algorithms by optimizing the framework that avoids vanishing gradient phenomena.

Our works also deal with an important question that only a few works address due to hardware limitations: how to ensure that a method's performance will scale with the size of the problems? The authors in \cite{bowles2024betterclassicalsubtleart} raises this issue effectively, by offering software tools to compare popular QML models with classical ones and observing that out-of-the-box classical machine learning models usually outperform the quantum classifiers for binary classification tasks. In addition, their results suggest that "quantumness" may not be the crucial ingredient for the small learning tasks considered.

By offering software tools that are tailored for Hamming-Weight preserving algorithms, and by mimicking the behavior of state-of-the-art classical deep-learning layers, we offer a solution that performs well in comparison with classical methods while offering an interesting running time advantage. Our software, that can be accessed through \cite{Monbroussou_Hamming_Weight_Preserving_2024}, allowed us to train our model on 10-label classification tasks that are far more complex than usual binary classification tasks used to illustrate QML methods and are commonly used in the classical Machine Learning literature.

We hope that our work, combined with other future studies on different figures of merits such as noise resilience, energy consumption, differential privacy, etc... will offer new perspectives for near term QML algorithms.
\section{Acknowledgment}

This work is supported by the H2020-FETOPEN Grant PHOQUSING (GA no.: 899544), the Engineering and Physical Sciences Research Council (grants EP/T001062/1), and the Naval Group Centre of Excellence for Information Human factors and Signature Management (CEMIS). ABG is supported by ANR JCJC TCS-NISQ ANR-22-CE47-0004, and by the PEPR integrated project EPiQ ANR-22-PETQ-0007 part of Plan France 2030.  This work is part of HQI initiative (www.hqi.fr) and is supported by France 2030 under the French National Research Agency award number ANR-22-PNCQ-0002. The authors warmly thank Manuel Rudolph for the helpful discussions.

\printbibliography %Prints bibliography

\begin{appendix}
    \section{Reminder on Hamming Weight Preserving circuits}\label{chap:HWPreserving_Resume}

In this Section, we present the main properties of Hamming Weight (HW) preserving gates. Let's define the basis of $n$-qubit states of HW $k$:
\begin{equation}\label{eq:HWBasis}
    B_k^n = \left\{ \ket{e} | \; e \in \{0,1\}^n \textrm{ and } HW(e) = k\right\} \, \textrm{,}
\end{equation}
with $HW(e)$ the number of $1$ in the bitstring $e$. We call \textbf{Hamming weight preserving} a $n$-qubit quantum circuit such that its corresponding unitary matrix $U$ follows:
\begin{equation}
    \forall k \in [n], \, \forall \ket{e} \in B_k^n, \quad U \cdot \ket{e} \in \textrm{span}(B_k^n) \, \textrm{.}
\end{equation}
Such a unitary is subspace preserving, and we can re-order its indexes to form a block diagonal matrix where each block is a unitary matrix corresponding to the action of the circuit in a specific subspace. 

\begin{figure}[H]
    \centering
    \includegraphics[width=0.75\textwidth]{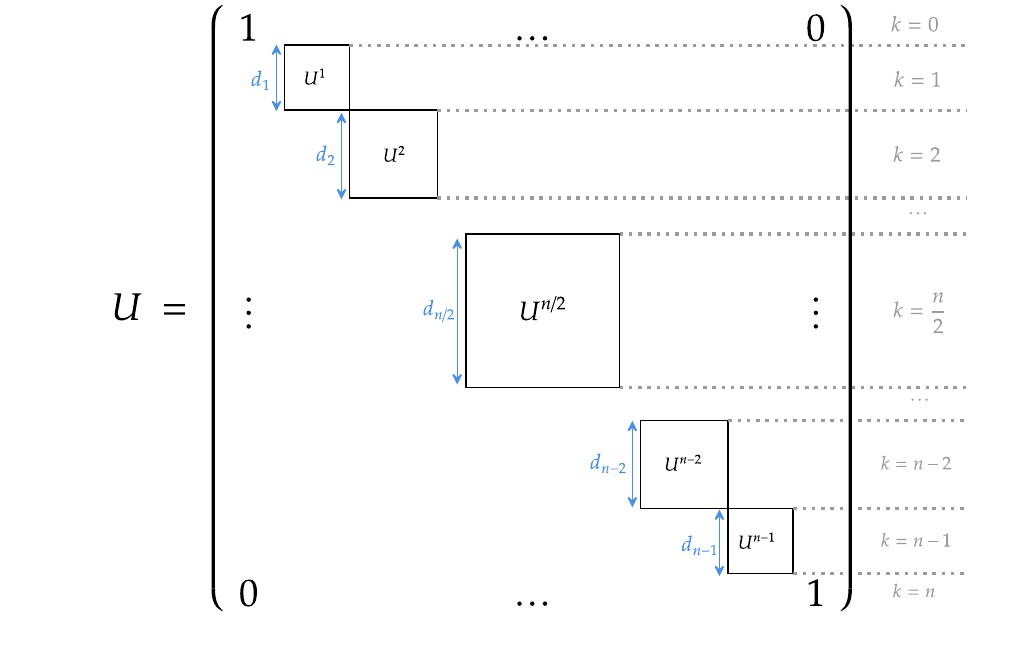}
    \caption{Block representation of the HW-preserving unitaries. $U$ is the $2^n \times 2^n$ unitary corresponding to a n-qubit HW-preserving quantum circuit. Each block $k$ is the unitary matrix corresponding to the preserved subspace of HW $k$, and the state basis $B_k^n$. Their size are $d_k\times d_k$ where $d_k = \binom{n}{k}$.}
    \label{fig:RBS_circuit_block_unitary}
\end{figure}

The most common HW preserving gate is the Reconfigurable Beam Splitter (RBS). This is the only HW preserving gate used in this article, and it is easy to implement or native on many quantum devices. RBS gates are presented through Definition \ref{def:RBSdef}. Notice that the definition of the RBS is the same as the one of the photonic Beam Splitter when considering a single photon. Theoretical guarantees on the training of RBS based circuit have been proposed by previous work \cite{monbroussou2023}, and it can be used to perform amplitude encoding on a specific subspace \cite{farias2024, Johri2020, monbroussou2023}.

\newpage
    \section{Hamming Weight Preserving Quantum Ansatz}\label{chap:HWPreserving_Ansatz}

In this Section, we recall the RBS-based ansatz presented in \cite{Cherrat2022}, that can reach any unitary in the subspace of HW $1$.

\begin{figure}[h]
    \centering
    \includegraphics[width=1.0\textwidth]{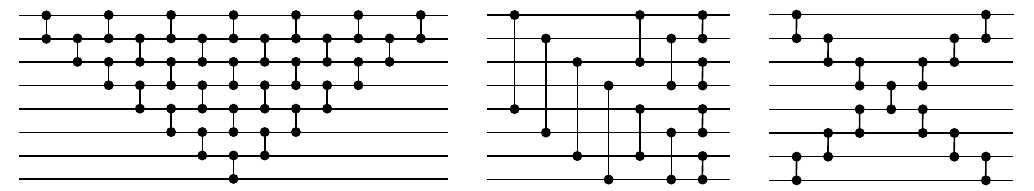}
    \caption{From left to right: Pyramid circuit, Butterfly Circuit, and X circuit. Vertical lines represent two-qubit RBS gates, parametrized with independent angles.}
    \label{fig:HW_Ansatz}
\end{figure}

Each $n$-qubit circuit presented in Figure \ref{fig:HW_Ansatz} is able to achieve any unitary matrix of dimension $n \times n$ in the subspace of HW $1$. Therefore, those circuit are good candidates to be applied on each register of HW $1$ in the quantum convolutional layers presented in Section \ref{subsec:Conv_layer}.
\end{appendix}

\end{document}